\documentclass[aps,prd,twocolumn,amsmath,nofootinbib,superscriptaddress]{revtex4-2}

\usepackage{graphicx}
\usepackage{float}
\usepackage{epstopdf}
\usepackage{epsfig}
\usepackage{amsmath}
\usepackage{subcaption}
\usepackage{color}
\usepackage{amssymb}   
\usepackage{amsfonts}  
\usepackage{amsbsy}    
\usepackage{pstricks}
\usepackage{slashed}
\usepackage{dcolumn}
\usepackage{xparse}
\usepackage{bm}
\usepackage{hyperref}
\usepackage{soul}
\usepackage{bbm}
\usepackage{tikz-feynman}

\tikzfeynmanset{compat=1.1.0} 
\usepackage{tikz} 
\usetikzlibrary{shapes,arrows,positioning,automata,backgrounds,calc,er,patterns}
\usepackage{contour}

\usepackage{caption}

\newcommand{\al}{\alpha}
\newcommand{\be}{\beta}

\newcommand{\beq}{\begin{equation}}
\newcommand{\eeq}{\end{equation}}
\newcommand{\ba}{\begin{array}}
\newcommand{\ea}{\end{array}}
\newcommand{\bea}{\begin{align}}
\newcommand{\eea}{\end{align}}
\newcommand{\bi}{\begin{itemize}}
\newcommand{\ei}{\end{itemize}}
\newcommand{\ben}{\begin{enumerate}}
\newcommand{\een}{\end{enumerate}}
\newcommand{\bc}{\begin{center}}
\newcommand{\ec}{\end{center}}
\newcommand{\bl}{\begin{flushleft}}
\newcommand{\el}{\end{flushleft}}
\newcommand{\br}{\begin{flushright}}
\newcommand{\er}{\end{flushright}}

\newcommand{\nn}{\nonumber \\}
\newcommand\Eqn[1]{Eq.~(\ref{#1})}  
\newcommand\Fig[1]{Fig.~\ref{#1}} 

\newcommand{\ie}{{i.e.}}

\newcommand{\fm}{{\rm fm}}

\newcommand{\GeV}{{\rm GeV}}

\newcommand\comment[1]{ \hbox{[{\it Comment suppressed here.}\/]} }
\newcommand\hide[1]{}
\newcommand{\skipover}[1]{}

\begin{document}
\captionsetup{justification=raggedright}
\title{A glimpse into pion gravitational form factor}

\author{Zanbin Xing}\email{xingzb@mail.nankai.edu.cn}
\affiliation{School of Physics, Nankai University, Tianjin 300071, China}
\author{Minghui Ding}\email{m.ding@hzdr.de}
\affiliation{Helmholtz-Zentrum Dresden-Rossendorf, Bautzner Landstra{\ss}e 400, 01328 Dresden, Germany}
\author{Lei Chang}\email{leichang@nankai.edu.cn}
\affiliation{School of Physics, Nankai University, Tianjin 300071, China}

\date{\today}
\begin{abstract}
We provide a novel approach to calculate the gravitational form factor of pion under the ladder approximation of the Bethe-Salpeter equation, with contact interactions. Central to this approach is a symmetry-preserving treatment of the dressed $\pi\pi$ amplitude, which shows explicitly the contributions from intrinsic quarks and bound states, the latter being necessary to produce the $D$-term of pion in the soft-pion limit. The approach we provide in this work can be applied to many processes of physical significance.   
\end{abstract}
\maketitle
\section{introduction}
The coupling of graviton to hadron via energy-momentum tensor (EMT) provides the gravitational form factor (GFF) of the hadron~\cite{Pagels:1966zza,Novikov:1980fa}, a form factor that has been of interest because fundamental physical quantities of hadrons, such as mass and spin, can be extracted from it. In addition to mass and spin, another fundamental physical quantity can be extracted from the GFF, the so called $D$-term~\cite{Polyakov:2018zvc}, which is related to the variation of the spatial components of the spacetime metric. Although all hadrons have a $D$-term, in most cases there is no fundamental principle governing their values.

The value of the $D$-term is unambiguously constrained in some special cases. For example, the Nambu-Goldstone boson for chiral symmetry breaking in  Quantum chromodynamics (QCD), pion, the $D$-term of this Goldstone boson is very special  in the soft-pion limit~\cite{Novikov:1980fa,Polyakov:1998ze}, and has the same value as that of the free spin-$0$ field~\cite{Pagels:1966zza,Hudson:2017xug}. Additionally, it has been pointed out that the slope of $D$-term correlates with chiral symmetry breaking~\cite{Leutwyler:1989tn}, and so the study of the $D$-term of pion may provide us with an opportunity to investigate the mechanism of dynamical chiral symmetry breaking in QCD, furthermore, to study the mechanism of emergent hadron mass. 

Inspired by this, the GFF of pion has now been studied both experimentally and theoretically. In experiments, the GFF of pion can be extracted from the pion pair production process $\gamma^{\star}\gamma \to \pi\pi$~\cite{Kumano:2017lhr}. In theory, there are relevant calculations for the pion GFF using phenomenological models, such as the Nambu--Jona-Lasinio (NJL)~\cite{Freese:2019bhb} and the chiral quark model~\cite{Broniowski:2008hx,Son:2014sna}. Meanwhile, the Dyson-Schwinger equations (DSEs) approach~\cite{Roberts:2021nhw} provides a continuum field theory approach to understanding hadrons, particularly for pion subject to chiral symmetry and its breaking constraints, where the relative uncertainties introduced by modeling and truncation can be well controlled in studies using this approach~\cite{Chang:2009zb,Xu:2022kng}. Given this advantage, the electromagnetic form factor~\cite{Chang:2013nia}, the distribution amplitude/function~\cite{Chang:2013pq,Ding:2019qlr} and the generalized parton distribution (GPD)~\cite{Raya:2021zrz} of pion have been calculated in the framework of this approach. In view of this, the study of the GFF seems to be very straightforward and in high demand. Therefore, in the following we will provide a general method to the calculation of the pion GFF using the DSEs approach. For simplicity, we will perform a symmetry-preserving regularized contact model~\cite{Xing:2022jtt} to illustrate the main ideas.

The paper is organized as follows. Sec.~\ref{sec::gff} introduces the general equation for the quark-graviton vertex, and relates the GFF to the $\pi\pi$ amplitude. Sec.~\ref{sec::pipiamplitude} discusses the solution for the $\pi\pi$ amplitude in the contact model and shows the contributions of scalar and vector propagators. Sec.~\ref{sec::results} provides numerical results, and the final section gives a summary.

\section{gravitational form factor}\label{sec::gff}

The gravitational form factors of pion can be extracted from its energy-momentum tensor (EMT), which can be expressed as follows:\footnote{We employ an Euclidean metric with $\{\gamma_\mu,\gamma_\nu\} = 2\delta_{\mu\nu}$; $\gamma_\mu^\dagger = \gamma_\mu$; $\gamma_5= \gamma_4\gamma_1\gamma_2\gamma_3$; and $a \cdot b = \sum_{i}^{4} a_i b_i$. The isospin symmetry in considered this work.}
\begin{align}
M^{\pi}_{\mu\nu}(Q^2)= &2P_\mu P_\nu A(Q^2)+\frac{1}{2}\left(Q_\mu Q_\nu-Q^2\delta_{\mu\nu}\right)D(Q^2)\nn
&+2m_\pi^2\delta_{\mu\nu}\bar{c}(Q^2)\,,
\end{align}
where $Q$ is the momentum transfer between the initial and final states, $P$ is the average momentum, and $m_\pi$ is the pion mass. Here we define $A(Q^2)=\sum_{a}A^{a}(Q^2)$, where $a$ denotes the different kinds of partons, and similarly for $D(Q^2)$ and $\bar{c}(Q^2)$. For hadrons with different spins, translational invariance requires the form factor $A(0)=1$ and EMT conservation implies $\bar{c}(Q^2)=0$. A spin-$0$ boson possesses an intrinsic $D$-term, defined as $D\equiv D(0)$. Specially, the Goldstone boson in the soft-pion limit, which has $D=-1$, see Ref.~\cite{Novikov:1980fa,Polyakov:1998ze}.

At the typical hadronic scale, the pion structure is dominated by dressed quarks with gluons hidden in these dressed quasiparticles~\cite{Cui:2022bxn}. We will stick to such a physical picture and consider only the quark part of the pion GFF. In the impulse approximation, the quark part EMT is expressed by a triangle diagram, as follows:
\begin{align}\label{eqn::iaemt}
\mathcal{M}^{\pi}_{\mu\nu}(Q^2)=&2N_c\text{tr}\int_q \Gamma^{G}_{\mu\nu}(q+k,q+p)S(q+p)\nn
&\times i\Gamma_{\pi}(p)S(q)i\Gamma_{\pi}(-k)S(q+k)\,,
\end{align}
where $p=P-Q/2$, $k=P+Q/2$ is the total momentum of the incoming and outgoing pion, respectively, the factor $2$ comes from the isospin symmetry of the $u,d$ quarks, $N_c=3$ is the color degree of freedom and the trace is over Dirac space, $\int_q$ represents a four-dimensional integral. Two building blocks in \Eqn{eqn::iaemt} that have been intensively studied are the dressed quark propagator $S(q)$ and the pion Bethe-Salpeter amplitude (BSA) $\Gamma_{\pi}(P)$, and here we neglect the relative momentum of the pion BSA in the notation for simplicity. They follow the corresponding quark gap equation and the Bethe-Salpeter equation (BSE), which are consistently truncated under the constraints of the Ward-Takahashi identities (WTI)~\cite{Chang:2009zb}. However, the third building block, the quark-graviton vertex (QGV) $\Gamma^{G}_{\mu\nu}(k^{+},k^{-})$, is a brand new object, satisfying the following corresponding gravitational Bethe-Salpeter equation (GBSE), 
\begin{align}\label{eqn::lgbse}
\Gamma^{G}_{\mu\nu}(k^{+},k^{-})=&\gamma^{G}_{\mu\nu}(k^{+},k^{-})\nn
&-\frac{4}{3}\int_q \mathcal{G}_{\al\be}(k-q)\gamma_{\al}\chi^{G}_{\mu\nu}(q^{+},q^{-})\gamma_{\be}\,,
\end{align}
where $k^{+}=k+\eta Q$ and $k^{-}=k-(1-\eta)Q$ is the momentum of the outgoing and incoming quarks, respectively,  with an arbitrary momentum partition $\eta$. The ladder approximation has been applied in \Eqn{eqn::lgbse}. $\mathcal{G}_{\al\be}(k-q)$ is the effective gluon propagator, $\chi^{G}_{\mu\nu}(q^{+},q^{-})=S(q^{+})\Gamma^{G}_{\mu\nu}(q^{+},q^{-})S(q^{-})$ is the QGV Bethe-Salpeter wave function. $\gamma^{G}_{\mu\nu}(k^{+},k^{-})$ is the bare QGV of the form
\begin{align}\label{eqn::bareqgv}
\gamma^{G}_{\mu\nu}(k^+,k^-)=i\gamma_\mu\frac{k^+_\nu+k^-_\nu}{2}-\delta_{\mu\nu}\frac{S_{0}^{-1}(k^+)+S_{0}^{-1}(k^-)}{2}\,,
\end{align}
where $S^{-1}_{0}(k)=i\slashed{k}+m$ is the inverse of the bare quark propagator, with $m$ being the current quark mass.

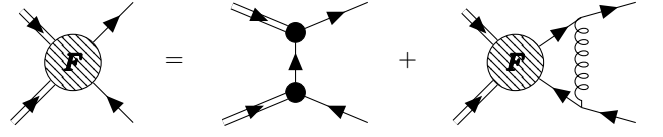
\begin{figure}
\centering
\begin{equation*}
\vcenter{\hbox{
\begin{tikzpicture}[scale=0.8]
\begin{feynman}
\vertex (a) at (-1,0) {}; 
\vertex [blob] (c) at (0,-1) {\contour{black}{$F$}};
\vertex (d) at (-1,-2);
\vertex (e) at (1,0);
\vertex (f) at (1,-2);

\diagram*{
(a) -- [double distance=2pt, with arrow=0.5] (c),
(d) -- [double distance=2pt, with arrow=0.5] (c),
(c) --  [fermion, edge label] (e),
(f) -- [fermion, edge label]  (c),
};
\end{feynman}
\end{tikzpicture}}}
\quad = \quad
\vcenter{\hbox{\begin{tikzpicture}[scale=0.8]
\begin{feynman}
\vertex (a) at (-1.2,0) {}; 
\vertex [dot] (b) at (0,-0.5) {1};
\vertex [dot] (c) at (0,-1.5) {1};
\vertex (d) at (-1.2,-2);
\vertex (e) at (1.2,0);
\vertex (f) at (1.2,-2);

\diagram*{
(a) -- [double distance=2pt, with arrow=0.5] (b),
(c) -- [fermion] (b),
(d) -- [double distance=2pt, with arrow=0.5] (c),
(b) --  [fermion] (e),
(f) -- [fermion]  (c),
};
\end{feynman}
\end{tikzpicture}}}
\quad + \quad
\vcenter{\hbox{\begin{tikzpicture}[scale=0.8]
\begin{feynman}
\vertex (a) at (-1,0) {}; 
\vertex [blob] (c) at (0,-1) {\contour{black}{$F$}};
\vertex (d) at (-1,-2);
\vertex (e) at (1.1,-0.25);
\vertex (f) at (1.1,-1.75);
\vertex (g) at (2,0);
\vertex (h) at (2,-2);

\diagram*{
(a) -- [double distance=2pt, with arrow=0.5] (c),
(d) -- [double distance=2pt, with arrow=0.5] (c),
(c) --  [fermion] (e),
(f) -- [fermion]  (c),
(e) --  [fermion] (g),
(h) -- [fermion] (f) ,
(e) -- [gluon]  (f),
};
\end{feynman}
\end{tikzpicture}}}
\end{equation*}
\caption{
BSE for the dressed $\pi\pi$ amplitude in graphical form. Solid and coiled lines represent quarks and gluon,  respectively. Double-solid lines are pions, blobs are dressed $\pi\pi$ amplitudes and filled circles are pion BSAs.}
\label{fig::pipiamplitude}
\end{figure}

The traditional way to compute the triangle diagram is to first solve the fully dressed QGV in \Eqn{eqn::lgbse} and then insert it into \Eqn{eqn::iaemt}. However, due to the complexity of the  dressed QGV, we decide not to compute the dressed QGV directly. Instead, we propose a novel method to compute the GFF in the impulse approximation, based on the connection between the the ladder approximation and the impulse approximation in \Eqn{eqn::iaemt}. The quark part EMT can be expressed as
\begin{align}\label{eqn::emt}
\mathcal{M}^{\pi}_{\mu\nu}(Q^2)=&2N_c\text{tr}\int_q \gamma^{G}_{\mu\nu}(q+k,q+p) S(q+p)\nn
&\times i^2F(q,p,-k) S(q+k)\,,
\end{align}
where $F(k,P_1,P_2)$ is the dressed amputated $\pi\pi$ amplitude we introduce, which satisfies its corresponding BSE in the ladder approximation, 
\begin{eqnarray}\label{eqn::pipiamplitude}
&&F(k,P_1,P_2)=F_0(k,P_1,P_2)+\Sigma^F(k,P_1,P_2)\,,
\end{eqnarray}
with the bare $\pi\pi$ amplitude
\begin{eqnarray}
F_0(k,P_1,P_2)=\Gamma_{\pi}(P_1)S(k)\Gamma_{\pi}(P_2)\,,
\end{eqnarray}
and the self-energy term
\begin{align}\label{eqn::sigma}
\Sigma^F(k,P_1,P_2)=&-\frac{4}{3}\int_q \mathcal{G}_{\al\be}(k-q)\gamma_{\al}S(q+P_1)\nn
&\times F(q,P_1,P_2)S(q-P_2)\gamma_{\be}\,.
\end{align}
The graphical presentation of \Eqn{eqn::pipiamplitude} is shown in \Fig{fig::pipiamplitude}.

It follows that the triangle diagram can be calculated in two ways, either by solving for the fully dressed QGV or by solving for the fully dressed $\pi\pi$ amplitude. The equivalence of these two ways of describing the triangle diagram was formally presented in Ref.~\cite{Bando:1993qy}. The $\pi\pi$ amplitude we introduced, has been calculated within rainbow-ladder truncation of DSEs to sketch the $\pi-\pi$ scattering process~\cite{Cotanch:2002vj}. It is worth noting that, for $\pi-\pi$ scattering processes, the calculation of $\pi\pi$ amplitude is necessary to ensure symmetry. By contrast, for the triangle diagram in \Eqn{eqn::iaemt} under the impulse approximation, since it is self-consistent with the ladder approximation, if the fully dressed QGV in \Eqn{eqn::lgbse} is known, there is in principle no need to perform $\pi\pi$ amplitude calculations. Yet, as we mentioned earlier, a fully dressed QGV is very complex. Instead, as one will see in the next section, the dressed $\pi\pi$ amplitude can be formally solved in the contact model, and one will read a lot of interesting physics directly from its solution.  The GFF can then be systematically obtained, in the absence of any information about the dressed QGV.

\section{dressed $\pi\pi$ amplitude in the contact model}\label{sec::pipiamplitude}
In the contact model~\cite{Gutierrez-Guerrero:2010waf}, the effective gluon propagator in \Eqn{eqn::lgbse} is
\begin{align}
\mathcal{G}_{\al\be}(k-q)=\frac{\delta_{\al\be}}{m_g^2}\,,
\end{align}
where $m_g$ is a gluon mass scale. Consequently, the self-energy term in \Eqn{eqn::sigma} becomes
\begin{align}
\Sigma^F(P_1,P_2)=-\Delta_{g}\int_q \gamma_{\al}S(q_1)F(q,P_1,P_2)S(q_2)\gamma_{\al}\,,
\end{align}
where we define for convenience $\Delta_{g}=\frac{4}{3m_g^2}$, $q_1=q+P_1$, and $q_2=q-P_2$. Note that $\Sigma^F(P_1,P_2)$ is now independent of the quark momentum $k$, which is its feature in the contact interaction model.

The self-energy term $\Sigma^F(P_1,P_2)$ can be generally written as
\begin{align}
	\Sigma^F(P_1,P_2)=\sum^4_{i=1} t_iT_i\,,
\end{align}
where $T_i$ is a set of orthogonal basis
\begin{eqnarray}\label{eqn::basis}
&&T_1=\mathbbm{1}\,,\nn
&&T_2=\frac{-i}{M}\slashed{K}\,,\nn
&&T_3=\frac{-i}{M}\slashed{Z}\,,\nn
&&T_4=\frac{i}{M^2}\sigma_{\mu\nu}Z_\mu K_\nu\,,
\end{eqnarray}
with $K=(P_1-P_2)/2$, $Z=-(P_1+P_2)$, and $M$ is the quark mass function, which is momentum independent in the contact model. The Kinematic relations entail $K\cdot Z=0$, $K^2=-m_\pi^2-Z^2/4$, thus the dressing scalar functions $t_i$ are functions of $Z^2$, \ie, $t_i=t_i(Z^2)$.  The dependence of the dressing scalar functions on the quark mass function $M$ and pion mass $m_\pi$ is suppressed for simplicity. Consequently, the BSE of the dressed $\pi\pi$ amplitude is equivalent to the equation satisfied by the self-energy, \ie,
\begin{align}\label{eqn::sigmaci}
\Sigma^F(P_1,P_2)=&\Sigma^{F_0}(P_1,P_2)\nn
&-\Delta_{g}\int_q \gamma_{\al}S(q_1)\Sigma^F(P_1,P_2)S(q_2)\gamma_{\al}\,,
\end{align}
where the inhomogeneous term is
\begin{align}
\Sigma^{F_0}(P_1,P_2)=-\Delta_{g}\int_q \gamma_{\al}S(q_1)F_{0}(q,P_1,P_2)S(q_2)\gamma_{\al}\,,
\end{align}
which can obviously also be decomposed in terms of the orthogonal basis in \Eqn{eqn::basis}, \ie, $\Sigma^{F_0}(P_1,P_2)=\sum^4_{i=1} b_{i}T_i$.

Considering \Eqn{eqn::sigmaci}, the charge conjugate symmetry requires the structure $T_3$ to vanish. Moreover, the vector-vector contact interaction can not give rise to the tensor-like structure $T_4$. Given this, only two orthogonal structures $T_{1,2}$ remain. Projecting \Eqn{eqn::sigmaci} onto $T_{1,2}$, one can obtain the corresponding dressing scalar functions $t_{1,2}$. By doing so, it is found that \Eqn{eqn::sigmaci} is decoupled into two equations, and their solutions are simple as
\begin{eqnarray}
t_1(Z^2)&=&\frac{b_{1}(Z^2)}{1+\Delta_{g}f_{s}(Z^2)}\,,\nn
t_2(Z^2)&=&\frac{b_{2}(Z^2)}{1+\Delta_{g}f_{v}(Z^2)}\,,
\end{eqnarray}
where the defined functions in the denominators are
\begin{eqnarray}
f_{s}(Z^2)&=&\text{tr}\int_q S(q_1) S(q_2)\,,\nn 
f_{v}(Z^2)&=&\frac{K_\mu K_\nu}{2K^2}\text{tr}\int_q i\gamma_{\mu}^TS(q_1) i\gamma_{\nu}^T S(q_2)\,,
\end{eqnarray}
with $\gamma_{\mu}^T=\gamma_{\mu}-\frac{Z_\mu\slashed{Z}}{Z^2}$. $f_{s}$ and $f_{v}$, as can be read directly from their expressions, correspond to the polarization of scalar and vector mesons, respectively. The defined functions in the numerators are
\begin{align}
b_{1}(Z^2)=&-\frac{1}{4}\Delta_{g}\text{tr}\int_q T_1\gamma_{\al}S(q_1)F_0(q,P_1,P_2)S(q_2)\gamma_{\al}\,,\nn
b_{2}(Z^2)=&\frac{M^2}{4K^2}\Delta_{g} \text{tr}\int_q T_2\gamma_{\al}S(q_1)F_0(q,P_1,P_2)S(q_2)\gamma_{\al}\,.
\end{align}
Recalling that the definition of the pion scalar form factor $F_s$ and the electromagnetic form factor $F_{\text{em}}$ in the impulse approximation are
\begin{align}\label{eqn::scalarandemff}
iF_s(Z^2)=&N_c \text{tr}\int_q i\Gamma_{I}(Z)S(q_1)i^{2}F_0(q,P_1,P_2)S(q_2)\,,\nn
2K_{\mu}F_{\text{em}}(Z^2)=&N_c \text{tr}\int_q i\Gamma_{\mu}(Z)S(q_1)i^{2}F_0(q,P_1,P_2)S(q_2)\,,
\end{align}
with $\Gamma_{I}(Z)$ being the dressed quark-scalar vertex and $\Gamma_{\mu}(Z)$ being the dressed quark-photon vertex, one immediately realizes that 
\begin{align}
N_c b_{1}(Z^2)=&\Delta_{g} F^{b}_s(Z^2)\,,\nn
N_c b_{2}(Z^2)=&-M \Delta_{g} F^{b}_{\text{em}}(Z^2)\,,
\end{align}
where $F^{b}_s(Z^2)$ and $F^{b}_{\text{em}}(Z^2)$ are computed by replacing the dressed vertices with the bare vertices, \ie,  $\Gamma_{I}\to\mathbbm{1}$ and $\Gamma_{\mu}\to\gamma_\mu$. Therefore, $b_1$ and $b_2$ are connected to the scalar and electromagnetic form factors respectively.

Substituting the results for $t_{1,2}$ into the general expression for the self-energy term, we obtain the solution for the self-energy in \Eqn{eqn::sigmaci} as 
\begin{align}\label{eqn::sigmasolu}
N_c\Sigma^F(P_1,P_2)=F^{b}_s(Z^2)\Delta_{s}(Z^2)+i\slashed{K}F^{b}_{\text{em}}(Z^2)\Delta_{v}(Z^2)\,,
\end{align}
where $\Delta_{s}$ and $\Delta_{v}$ are given by
\begin{eqnarray}
&&\Delta_{s}(Z^2)=\frac{1}{\Delta_{g}^{-1}+f_s(Z^2)}\,,\nn
&&\Delta_{v}(Z^2)=\frac{1}{\Delta_{g}^{-1}+f_v(Z^2)}\,.
\end{eqnarray}
The poles of the scalar and vector mesons are shown in the expressions for $\Delta_{s}$ and $\Delta_{v}$. These two expressions can also be considered as part of, up to the Lorentz structure, of the scalar and vector meson propagators.
After obtaining the self-energy solution, the solution for the dressed $\pi\pi$ amplitude can eventually be written as
\begin{eqnarray}\label{eqn::GFFa}
F(k,P_1,P_2)=F_0(k,P_1,P_2)+\Sigma^F(P_1,P_2)\,.
\end{eqnarray}

The explicit form of $\Sigma^F(P_1,P_2)$ in \Eqn{eqn::sigmasolu} shows rich physics, including not only the  scalar and vector poles, but also the scalar and vector form factors. Consequently, the same is true of the $\pi\pi$ amplitude in \Eqn{eqn::GFFa}. As a result, the bare vertex can then be used to extract the GFF from the $\pi\pi$ amplitude. Additionally, it is worth being pointed out that such direct connections between $\Sigma^F(P_1,P_2)$ and the form factors in \Eqn{eqn::sigmasolu} are obtained due to the momentum-independent gluon propagator in the contact model. Nonetheless, even with the momentum-dependent gluon propagator, there must be implicit connections between the form factor and the $\pi\pi$ amplitude. Further investigation of the $\pi\pi$ amplitude with QCD-based interactions is on the agenda.

\section{GFF results}\label{sec::results}
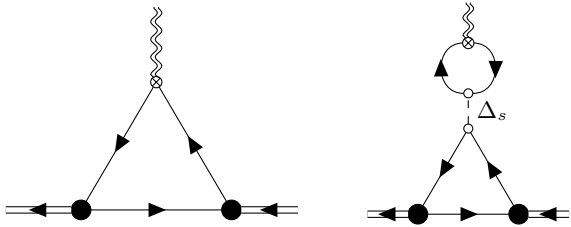
\begin{figure}
\centering
\begin{equation*}
\vcenter{\hbox{\begin{tikzpicture}[scale=1.0]
\begin{feynman}
\vertex (a) at (0,-1.2) {}; 
\vertex [dot] (b) at (-1,-1.2) {1};
\vertex [dot] (c) at (-3,-1.2) {1};
\vertex (d) at (-4,-1.2);
\vertex [scale=0.5, crossed dot](e) at (-2,0.5) {};
\vertex (f) at (-1.97,0.56);
\vertex (g) at (-1.97,1.5);
\vertex (h) at (-2.03,0.56);
\vertex (i) at (-2.03,1.5);

\diagram*{
(a) -- [double distance=2pt, with arrow=0.5] (b),
(c) -- [fermion] (b) --  [fermion] (e)--  [fermion] (c),
(c) -- [double distance=2pt, with arrow=0.5] (d),
(g) -- [boson]  (f) -- (e),
(i) -- [boson]  (h) -- (e),
};
\end{feynman}
\end{tikzpicture}}}
\quad   \quad
\vcenter{\hbox{\begin{tikzpicture}[scale=0.67]
\begin{feynman}
\vertex (a) at (0,-1.2); 
\vertex [dot] (b) at (-1,-1.2) {1};
\vertex [dot] (c) at (-3,-1.2) {1};
\vertex (d) at (-4,-1.2);
\vertex [scale=0.8, empty dot] (e) at (-2,0.5) {};
\vertex [scale=0.8, empty dot] (f) at (-2,1.2) {};
\vertex [scale=0.5, crossed dot] (g) at (-2,2.2) {};
\vertex (h) at (-1.96,2.29);
\vertex (i) at (-1.96,3);
\vertex (j) at (-2.04,2.29);
\vertex (k) at (-2.04,3);

\diagram*{
(a) -- [double distance=2pt, with arrow=0.5] (b),
(c) -- [fermion] (b) --  [fermion] (e)--  [fermion] (c),
(c) -- [double distance=2pt, with arrow=0.5] (d),
(e) -- [scalar, edge label'=$\Delta_s$ ]  (f),
(f) -- [fermion, half left]  (g) -- [fermion, half left]  (f),
(i) -- [boson]  (h) -- (g),
(k) -- [boson]  (j) -- (g),
};
\end{feynman}
\end{tikzpicture}}}
\end{equation*}

\caption{\emph{Left:} $\mathcal{M}^{\pi}_{1\mu\nu}$, the impulse approximation with the bare QGV. \emph{Right:} $\mathcal{M}^{\pi}_{2\mu\nu}$, showing the scalar meson pole. Solid, double-solid, double-wavy and dashed lines represent quark, pion, graviton, and the effective scalar propagator $\Delta_s$, respectively. Filled, crossed and empty circles represent pion BSA, bare QGV $\gamma^{G}_{\mu\nu}$, and bare quark-scalar vertex, respectively.}
\label{fig::feyngff}
\end{figure}

Substituting the dressed $\pi\pi$ amplitude in \Eqn{eqn::GFFa} into the quark part EMT in \Eqn{eqn::emt}, the EMT of pion can be expressed as two parts,
\begin{eqnarray}
\mathcal{M}^{\pi}_{\mu\nu}(Q^2)&=&\mathcal{M}^{\pi}_{1\mu\nu}(Q^2)+\mathcal{M}^{\pi}_{2\mu\nu}(Q^2)\,,
\end{eqnarray}
with
\begin{align}
\mathcal{M}^{\pi}_{1\mu\nu}=&2N_c\text{tr}\int_q \gamma^{G}_{\mu\nu}(q+k,q+p)S(q+p)\nn
&\times i^2F_0(q,p,-k)S(q+k)\,,\nn
\mathcal{M}^{\pi}_{2\mu\nu}=&2N_c\text{tr}\int_q \gamma^{G}_{\mu\nu}(q+k,q+p)S(q+p)\nn
&\times i^2\Sigma^{F}(p,-k)S(q+k)\,.
\end{align}
Here, $\mathcal{M}^{\pi}_{1\mu\nu}$ corresponds to the so-called impulse approximation with insertion of the bare QGV, $\gamma^{G}_{\mu\nu}$, while $\mathcal{M}^{\pi}_{2\mu\nu}$ provides the contribution from the scalar bound state, and graphical representations of both are shown in \Fig{fig::feyngff}. It should be noted that the vector bound states do not contribute to the pion GFF. More notably, the appearance of $\mathcal{M}^{\pi}_{1\mu\nu}$ and $\mathcal{M}^{\pi}_{2\mu\nu}$ explicitly shows contributions from the intrinsic quarks and bound states.

\begin{figure}[t]
\includegraphics[width=8.6cm]{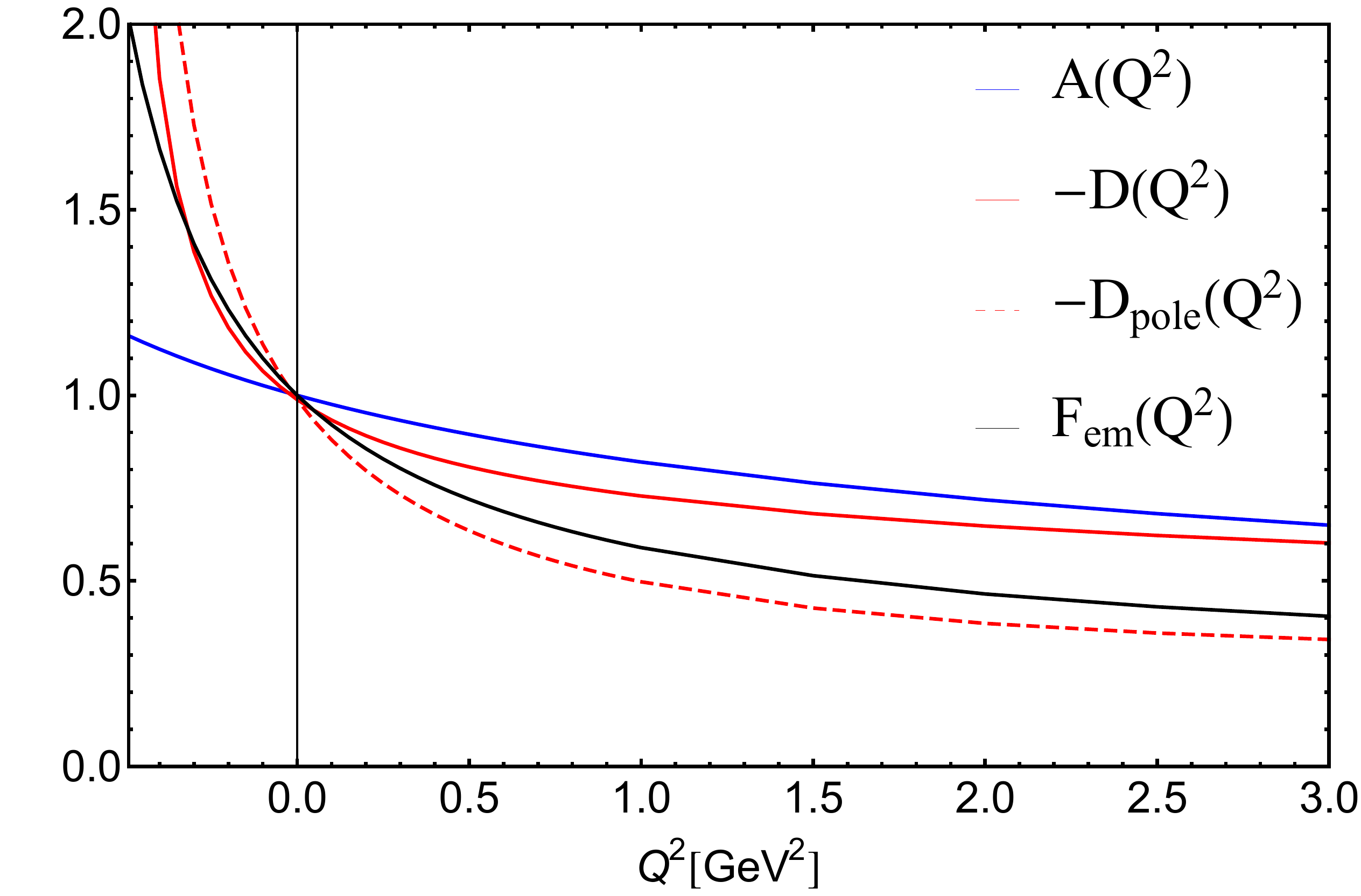}
\caption{Gravitational form factors $A(Q^2)$, $D(Q^2)$ and the electromagnetic form factor $F_{\text{em}}(Q^2)$. $D_{\text{pole}}(Q^2)$ is computed with the scalar-meson dominated Ansatz \Eqn{eqn::scalarpole}.}
\label{fig::gff}
\end{figure}

In order to calculate the $\pi\pi$ amplitude, and immediately afterwards the GFF, the dressed quark propagator and the pion BSA are required to be calculated. The general structures of the quark propagator and the pion BSA in the contact model are as follows
\begin{align}
S^{-1}(k)=&i\slashed{k}+M\,,\nn
\Gamma_{\pi}(P)=&i\gamma_5 E_\pi+\frac{\gamma_5\slashed{P}}{M}F_\pi\,.
\end{align}
The procedures for calculating these two quantities are described in Ref.~\cite{Gutierrez-Guerrero:2010waf}. In the course of the calculations, we use the regularization approach and  parameters developed in Ref.~\cite{Xing:2022jtt}. The results of all the required values are listed in the Table~\ref{tab:quarkandpion}. By substituting these numerical results into the expression of the quark part ETM, we obtain directly the numerical results of the form factors, whose $Q^{2}$ dependencies are shown in \Fig{fig::gff}.

Remarkably, we find that the form factor $A(Q^{2})$ has only the contribution from $\mathcal{M}^{\pi}_{1\mu\nu}$, which does not have any singularities. In contrast, the form factor $D(Q^{2})$ and the electromagnetic form factor $F_{\text{em}}(Q^{2})$ contain the scalar meson and vector meson poles in the time-like region, respectively, with partial contributions from $\mathcal{M}^{\pi}_{2\mu\nu}$.

\begin{table}[t]
\caption{\label{tab:quarkandpion} Numerical results for the gap equation and pion BSE, with masses are given in $\GeV$ and BSAs as dimensionless.}
\begin{tabular*}{0.45\textwidth}{c @{\extracolsep{\fill}} ccc}
\hline
$M$ &$m_{\pi}$ & $E_{\pi}/\sqrt{2}$ & $F_{\pi}/\sqrt{2}$\\
\hline
0.368 &0.140 & 3.595 & 0.475\\
\hline
\end{tabular*}
\end{table}

In the chiral limit, we obtain the $D$-term where the contributions of $\mathcal{M}^{\pi}_{1\mu\nu}$ and $\mathcal{M}^{\pi}_{2\mu\nu}$ are
\begin{eqnarray}
D_1(0)=-\frac{E_{\pi}-6F_{\pi}}{3(E_{\pi}-2F_{\pi})}\,,\nn
D_2(0)=-\frac{2E_{\pi}}{3(E_{\pi}-2F_{\pi})}\,,
\end{eqnarray}
and the sum of the two gives the total contribution of the pion $D$-term as
\begin{equation}
D=D_1(0)+D_2(0)=-1\,.
\end{equation}
This is exactly the result predicted by the soft-pion theorem. We also note that in the limit of $F_{\pi}=0$, our results are consistent with those in the Ref.~\cite{Freese:2019bhb} applying the NJL method.
In addition to the impulse approximation, the second term $\mathcal{M}^{\pi}_{2\mu\nu}$ is necessary to obtain the correct $D$-term. We would like to highlight that this result is a consequence of the chiral symmetry-preserving truncation of the model. In $\mathcal{M}^{\pi}_{2\mu\nu}$, an effective scalar propagator is introduced which can be rewritten as
\begin{align}
\Delta_{s}(Q^2)=\frac{1}{\Delta_{g}^{-1}+f_s(Q^2)}=\Delta_{g}\sum_{n}(-\Delta_{g})^{n}\left[f_s(Q^2)\right]^{n}\,.
\end{align}
From the expansion form of this expression we can venture a guess that the second term $\mathcal{M}^{\pi}_{2\mu\nu}$ is likely to be similar to the contact Feynman diagram introduced in Ref.~\cite{Polyakov:1999gs}.

After observing the value of the form factor at $Q^2=0$, we can also observe the trend of the form factor as $Q^2$ changes and, more precisely, how fast it falls. The slope of the electromagnetic form factor is found to be greater than that of $-D(Q^2)$, contrary to the previous prediction in Ref.~\cite{Raya:2021zrz}. This is mainly because the pion scalar form factor is hard and not well described in the contact model, as explained in Ref.~\cite{Wang:2022mrh}. Therefore, complementary to the solved $\Delta_s(Q^2)$, we also use a monopole Ansatz
\begin{equation}\label{eqn::scalarpole}
\Delta_s^{\text{pole}}(Q^2)=\frac{\Delta_s(0)}{1+Q^2/m_\sigma^2}\,,
\end{equation}
with $m_\sigma$ being the $\sigma$ meson mass extracted from the scalar meson pole. In this way, the Ansatz manifests a scalar meson-dominated Ansatz. The form factor computed using this Ansatz, labelled $D_{\text{pole}}(Q^2)$ is depicted in \Fig{fig::gff} with a red dashed line. For a better illustration, we also calculate the radii of the corresponding form factors, defined as
\begin{equation}
\langle r_\#^2\rangle=-\frac{6}{F_\#(0)}\frac{\partial F_\#(Q^2)}{\partial Q^2}\,,
\end{equation}
and present the results in Table \ref{tab:radii}.
Here we use $\#$ to denote the different types of radius and form factor. The radius we obtain for $A(Q^{2})$ is found to be around half the electromagnetic radius, which is consistent with the calculation in the large $N_c$ limit~\cite{Polyakov:1998ze}. The radius of $D(Q^{2})$ without the monopole Ansatz lies between the radii of $A(Q^{2})$ and $F_{\text{em}}(Q^2)$.  The monopole Ansatz increases the radius of $D(Q^{2})$ to a large extent, and the ratio of the radius of $D_{\text{pole}}(Q^{2})$ obtained using the monopole Ansatz, to the electromagnetic radius, is $1.2$, close to the previous prediction in Ref.~\cite{Raya:2021zrz}. 
\begin{table}[t]
\caption{\label{tab:radii} Radii of the corresponding form factors, unit is $\fm$.}
\begin{tabular*}{0.45\textwidth}{c @{\extracolsep{\fill}} ccc}
\hline
$\langle r_{A}^2\rangle^{1/2}$ &$\langle r_{D}^2\rangle^{1/2}$ &$\langle r_{D_{\text{pole}}}^2\rangle^{1/2}$ & $\langle r_{\text{em}}^2\rangle^{1/2}$\\
\hline
0.24 &0.39 & 0.54 & 0.45\\
\hline
\end{tabular*}
\end{table}

Final remark: Another issue in our calculations is that $\bar{c}$ does not vanish. 
Contracting \Fig{fig::feyngff} by $Q_{\mu}$, and using charge conjugation and translational invariance, one obtains
\begin{align}\label{eqn::cbar}
&Q_{\mu}\mathcal{M}^{\pi}_{\mu\nu}(Q^2)
=2Q_{\nu}(M-m)F_{s}(Q^2)\nn
&-N_c\text{tr}\int_q i^2\Gamma_{\pi}(p)S(q)\Gamma_{\pi}(-k)
\frac{Q_{\nu}}{2}\left[S(q+k)+S(q+p)\right]\,,
\end{align}
where $F_{s}(Q^2)$, as defined in \Eqn{eqn::scalarandemff}, is the pion scalar form factor computed with a fully dressed quark-scalar vertex. It is obvious from this equation that $\bar{c}\neq0$. This issue has been overlooked in Ref.~\cite{Broniowski:2008hx,Son:2014sna}.
In particular, note that the first term on the right-hand side of \Eqn{eqn::cbar} is proportional to $M-m$, which vanishes if the quark $m$ of the current mass is replaced by the dressed quark mass $M$. And, since the bare QGV $\gamma^G_{\mu\nu}$ is the only element that explicitly contains $m$, this substitution is equivalent to modifying the bare QGV to  
\begin{equation}\label{eqn::tildegamma}
\tilde{\gamma}^G_{\mu\nu}=\gamma^G_{\mu\nu}+(m-M)\delta_{\mu\nu}\,.
\end{equation}
Indeed, Ref.~\cite{Freese:2019bhb}, which uses the NJL method, shows that using a five-point vertex, one can derive a driving term in the GBSE, precisely $(m-M)\delta_{\mu\nu}$ in \Eqn{eqn::tildegamma}. 
If in \Eqn{eqn::lgbse} using $\tilde{\gamma}^G_{\mu\nu}$ instead of $\gamma^G_{\mu\nu}$ as the inhomogeneous term, one finds that under the constraints of such an equation, $\Gamma^G_{\mu\nu}$ satisfies exactly the gravitational WTI in the framework of the contact interaction model~\cite{Brout:1966oea}. However, the presence of the second term on the right-hand side of \Eqn{eqn::cbar} implies that the impulse approximation is not sufficient to guarantee conservation of energy-momentum even if GBSE matches GWTI. 
 The missing terms, complements to \Fig{fig::feyngff} to recover $\bar{c}=0$, can be understood in terms of the fact that the graviton can interact with, quarks exchanged between any two gluons in the quark-antiquark scattering kernel. A similar situation arises when using the ladder approximation to calculate the pion distribution function, and Ref.~\cite{Chang:2014lva} proposed an additional diagram as a compensation to ensure momentum conservation. How to remedy this shortcoming of the ladder truncation method is beyond the scope of this article.

\section{summary}
In this paper, we provide a symmetry-preserving way to calculate the pion gravitational form factor under the ladder approximation of the BSEs approach. Instead of solving for the dressed quark-graviton vertex, as is commonly used in other form factor calculations with the impulse approximation, we have turned to expressing the form factor by considering the dressed $\pi\pi$ amplitude and coupling the dressed $\pi\pi$ amplitude to the corresponding bare vertex. In the practical calculations, a contact model is used and the $\pi\pi$ amplitude is found to contain additional scalar and vector bound state contributions. Using this symmetry-preserving approach, the result of the calculation, $A(0)=1$ reflects that the quark carries all of the pion momentum, as expected on the typical hadron scale. Moreover, the $D$-term $D=-1$ in the chiral limit reproduces exactly what would be expected in the soft-pion limit. Unfortunately, the ladder approximation does not guarantee energy-momentum conservation, which can be seen as a defect of the ladder approximation in its own right, independent of the gluon interaction model. 

Although we have chosen a simple contact model to illustrate the computational procedure of the GFF, the present approach can be extended to realistic cases using momentum-dependent interactions. Study using realistic gluon interactions is ongoing and its development is essential as it can reveal more features of the GFF, such as the order of the form factor radii and its large $Q^{2}$ behavior, thus further rescuing some of the shortcomings of using the contact model. We also note that the study of the contribution of the pion loop to the GFF is also urgent and should be addressed in the future. Finally, we would like to highlight that, in addition to its application to the GFF calculations here, the $\pi\pi$ amplitude is a desired ingredient for the calculation of the $m\to n$ (with $m+n=4$) process, such as the pion-pair production process $\gamma^{\star}\gamma \to \pi\pi$. Furthermore, the $\pi\pi$ amplitude can be extended to analogous amplitudes consisting of other mesons. These studies will be a useful addition to the current work.

\hspace*{\fill}\ 
\begin{acknowledgments}
L. Chang is grateful for constructive conversations with K. Raya. This work benefited from discussions and presentations at the workshop ``Revealing emergent mass through studies of hadron spectra and structure", hosted by $\text{ECT}^{\star}$, on 12-16 September 2022. Work supported by National Natural Science Foundation of China (grant no. 12135007). M. Ding is grateful for support by Helmholtz-Zentrum Dresden-Rossendorf High Potential Programme.
\end{acknowledgments}

\bibliography{apstemplateNotes}

\providecommand{\noopsort}[1]{}\providecommand{\singleletter}[1]{#1}%
\begin{thebibliography}{26}
\providecommand{\natexlab}[1]{#1}
\providecommand{\url}[1]{\texttt{#1}}
\expandafter\ifx\csname urlstyle\endcsname\relax
  \providecommand{\doi}[1]{doi: #1}\else
  \providecommand{\doi}{doi: \begingroup \urlstyle{rm}\Url}\fi

\bibitem[Pagels(1966)]{Pagels:1966zza}
Heinz Pagels.
\newblock {Energy-Momentum Structure Form Factors of Particles}.
\newblock \emph{Phys. Rev.}, 144:\penalty0 1250--1260, 1966.
\newblock \doi{10.1103/PhysRev.144.1250}.

\bibitem[Novikov and Shifman(1981)]{Novikov:1980fa}
V.~A. Novikov and Mikhail~A. Shifman.
\newblock {Comment on the psi-prime ---\ensuremath{>} J/psi pi pi Decay}.
\newblock \emph{Z. Phys. C}, 8:\penalty0 43, 1981.
\newblock \doi{10.1007/BF01429829}.

\bibitem[Polyakov and Schweitzer(2018)]{Polyakov:2018zvc}
Maxim~V. Polyakov and Peter Schweitzer.
\newblock {Forces inside hadrons: pressure, surface tension, mechanical radius,
  and all that}.
\newblock \emph{Int. J. Mod. Phys. A}, 33\penalty0 (26):\penalty0 1830025,
  2018.
\newblock \doi{10.1142/S0217751X18300259}.

\bibitem[Polyakov(1999)]{Polyakov:1998ze}
Maxim~V. Polyakov.
\newblock {Hard exclusive electroproduction of two pions and their resonances}.
\newblock \emph{Nucl. Phys. B}, 555:\penalty0 231, 1999.
\newblock \doi{10.1016/S0550-3213(99)00314-4}.

\bibitem[Hudson and Schweitzer(2017)]{Hudson:2017xug}
Jonathan Hudson and Peter Schweitzer.
\newblock {D term and the structure of pointlike and composed spin-0
  particles}.
\newblock \emph{Phys. Rev. D}, 96\penalty0 (11):\penalty0 114013, 2017.
\newblock \doi{10.1103/PhysRevD.96.114013}.

\bibitem[Leutwyler and Shifman(1989)]{Leutwyler:1989tn}
H.~Leutwyler and Mikhail~A. Shifman.
\newblock {GOLDSTONE BOSONS GENERATE PECULIAR CONFORMAL ANOMALIES}.
\newblock \emph{Phys. Lett. B}, 221:\penalty0 384--388, 1989.
\newblock \doi{10.1016/0370-2693(89)91730-9}.

\bibitem[Kumano et~al.(2018)Kumano, Song, and Teryaev]{Kumano:2017lhr}
S.~Kumano, Qin-Tao Song, and O.~V. Teryaev.
\newblock {Hadron tomography by generalized distribution amplitudes in
  pion-pair production process $\gamma^* \gamma \rightarrow \pi^0 \pi^0 $ and
  gravitational form factors for pion}.
\newblock \emph{Phys. Rev. D}, 97\penalty0 (1):\penalty0 014020, 2018.
\newblock \doi{10.1103/PhysRevD.97.014020}.

\bibitem[Freese et~al.(2019)Freese, Freese, Clo\"et, and
  Clo\"et]{Freese:2019bhb}
Adam Freese, Adam Freese, Ian~C. Clo\"et, and Ian~C. Clo\"et.
\newblock {Gravitational form factors of light mesons}.
\newblock \emph{Phys. Rev. C}, 100\penalty0 (1):\penalty0 015201, 2019.
\newblock \doi{10.1103/PhysRevC.100.015201}.
\newblock [Erratum: Phys.Rev.C 105, 059901 (2022)].

\bibitem[Broniowski and Ruiz~Arriola(2008)]{Broniowski:2008hx}
Wojciech Broniowski and Enrique Ruiz~Arriola.
\newblock {Gravitational and higher-order form factors of the pion in chiral
  quark models}.
\newblock \emph{Phys. Rev. D}, 78:\penalty0 094011, 2008.
\newblock \doi{10.1103/PhysRevD.78.094011}.

\bibitem[Son and Kim(2014)]{Son:2014sna}
Hyeon-Dong Son and Hyun-Chul Kim.
\newblock {Stability of the pion and the pattern of chiral symmetry breaking}.
\newblock \emph{Phys. Rev. D}, 90\penalty0 (11):\penalty0 111901, 2014.
\newblock \doi{10.1103/PhysRevD.90.111901}.

\bibitem[Roberts et~al.(2021)Roberts, Richards, Horn, and
  Chang]{Roberts:2021nhw}
Craig~D. Roberts, David~G. Richards, Tanja Horn, and Lei Chang.
\newblock {Insights into the emergence of mass from studies of pion and kaon
  structure}.
\newblock \emph{Prog. Part. Nucl. Phys.}, 120:\penalty0 103883, 2021.
\newblock \doi{10.1016/j.ppnp.2021.103883}.

\bibitem[Chang and Roberts(2009)]{Chang:2009zb}
Lei Chang and Craig~D. Roberts.
\newblock {Sketching the Bethe-Salpeter kernel}.
\newblock \emph{Phys. Rev. Lett.}, 103:\penalty0 081601, 2009.
\newblock \doi{10.1103/PhysRevLett.103.081601}.

\bibitem[Xu et~al.(2022)Xu, Yao, Qin, Cui, and Roberts]{Xu:2022kng}
Zhen-Ni Xu, Zhao-Qian Yao, Si-Xue Qin, Zhu-Fang Cui, and Craig~D. Roberts.
\newblock {Bethe-Salpeter kernel and properties of strange-quark mesons}.
\newblock 8 2022.

\bibitem[Chang et~al.(2013{\natexlab{a}})Chang, Clo\"et, Roberts, Schmidt, and
  Tandy]{Chang:2013nia}
L.~Chang, I.~C. Clo\"et, C.~D. Roberts, S.~M. Schmidt, and P.~C. Tandy.
\newblock {Pion electromagnetic form factor at spacelike momenta}.
\newblock \emph{Phys. Rev. Lett.}, 111\penalty0 (14):\penalty0 141802,
  2013{\natexlab{a}}.
\newblock \doi{10.1103/PhysRevLett.111.141802}.

\bibitem[Chang et~al.(2013{\natexlab{b}})Chang, Cloet, Cobos-Martinez, Roberts,
  Schmidt, and Tandy]{Chang:2013pq}
Lei Chang, I.~C. Cloet, J.~J. Cobos-Martinez, C.~D. Roberts, S.~M. Schmidt, and
  P.~C. Tandy.
\newblock {Imaging dynamical chiral symmetry breaking: pion wave function on
  the light front}.
\newblock \emph{Phys. Rev. Lett.}, 110\penalty0 (13):\penalty0 132001,
  2013{\natexlab{b}}.
\newblock \doi{10.1103/PhysRevLett.110.132001}.

\bibitem[Ding et~al.(2020)Ding, Raya, Binosi, Chang, Roberts, and
  Schmidt]{Ding:2019qlr}
Minghui Ding, Kh\'epani Raya, Daniele Binosi, Lei Chang, Craig~D Roberts, and
  Sebastian~M Schmidt.
\newblock {Drawing insights from pion parton distributions}.
\newblock \emph{Chin. Phys. C}, 44\penalty0 (3):\penalty0 031002, 2020.
\newblock \doi{10.1088/1674-1137/44/3/031002}.

\bibitem[Raya et~al.(2022)Raya, Cui, Chang, Morgado, Roberts, and
  Rodriguez-Quintero]{Raya:2021zrz}
Khepani Raya, Zhu-Fang Cui, Lei Chang, Jose-Manuel Morgado, Craig~D. Roberts,
  and Jose Rodriguez-Quintero.
\newblock {Revealing pion and kaon structure via generalised parton
  distributions *}.
\newblock \emph{Chin. Phys. C}, 46\penalty0 (1):\penalty0 013105, 2022.
\newblock \doi{10.1088/1674-1137/ac3071}.

\bibitem[Xing and Chang(2022)]{Xing:2022jtt}
Zanbin Xing and Lei Chang.
\newblock {A symmetry preserving contact interaction treatment of the kaon}.
\newblock 10 2022.

\bibitem[Cui et~al.(2022)Cui, Ding, Morgado, Raya, Binosi, Chang, De~Soto,
  Roberts, Rodr\'\i{}guez-Quintero, and Schmidt]{Cui:2022bxn}
Z.~F. Cui, Minghui Ding, J.~M. Morgado, K.~Raya, D.~Binosi, L.~Chang,
  F.~De~Soto, C.~D. Roberts, J.~Rodr\'\i{}guez-Quintero, and S.~M. Schmidt.
\newblock {Emergence of pion parton distributions}.
\newblock \emph{Phys. Rev. D}, 105\penalty0 (9):\penalty0 L091502, 2022.
\newblock \doi{10.1103/PhysRevD.105.L091502}.

\bibitem[Bando et~al.(1994)Bando, Harada, and Kugo]{Bando:1993qy}
Masako Bando, Masayasu Harada, and Taichiro Kugo.
\newblock {External gauge invariance and anomaly in BS vertices and bound
  states}.
\newblock \emph{Prog. Theor. Phys.}, 91:\penalty0 927--948, 1994.
\newblock \doi{10.1143/PTP.91.927}.

\bibitem[Cotanch and Maris(2002)]{Cotanch:2002vj}
Stephen~R. Cotanch and Pieter Maris.
\newblock {QCD based quark description of pi pi scattering up to the sigma and
  rho region}.
\newblock \emph{Phys. Rev. D}, 66:\penalty0 116010, 2002.
\newblock \doi{10.1103/PhysRevD.66.116010}.

\bibitem[Gutierrez-Guerrero et~al.(2010)Gutierrez-Guerrero, Bashir, Cloet, and
  Roberts]{Gutierrez-Guerrero:2010waf}
L.~X. Gutierrez-Guerrero, A.~Bashir, I.~C. Cloet, and C.~D. Roberts.
\newblock {Pion form factor from a contact interaction}.
\newblock \emph{Phys. Rev. C}, 81:\penalty0 065202, 2010.
\newblock \doi{10.1103/PhysRevC.81.065202}.

\bibitem[Polyakov and Weiss(1999)]{Polyakov:1999gs}
Maxim~V. Polyakov and C.~Weiss.
\newblock {Skewed and double distributions in pion and nucleon}.
\newblock \emph{Phys. Rev. D}, 60:\penalty0 114017, 1999.
\newblock \doi{10.1103/PhysRevD.60.114017}.

\bibitem[Wang et~al.(2022)Wang, Xing, Kang, Raya, and Chang]{Wang:2022mrh}
Xiaobin Wang, Zanbin Xing, Jiayin Kang, Kh\'epani Raya, and Lei Chang.
\newblock {Pion scalar, vector, and tensor form factors from a contact
  interaction}.
\newblock \emph{Phys. Rev. D}, 106\penalty0 (5):\penalty0 054016, 2022.
\newblock \doi{10.1103/PhysRevD.106.054016}.

\bibitem[Brout and Englert(1966)]{Brout:1966oea}
R.~Brout and F.~Englert.
\newblock {Gravitational Ward Identity and the Principle of Equivalence}.
\newblock \emph{Phys. Rev.}, 141\penalty0 (4):\penalty0 1231--1232, 1966.
\newblock \doi{10.1103/PhysRev.141.1231}.

\bibitem[Chang et~al.(2014)Chang, Mezrag, Moutarde, Roberts,
  Rodr\'\i{}guez-Quintero, and Tandy]{Chang:2014lva}
Lei Chang, C\'edric Mezrag, Herv\'e Moutarde, Craig~D. Roberts, Jose
  Rodr\'\i{}guez-Quintero, and Peter~C. Tandy.
\newblock {Basic features of the pion valence-quark distribution function}.
\newblock \emph{Phys. Lett. B}, 737:\penalty0 23--29, 2014.
\newblock \doi{10.1016/j.physletb.2014.08.009}.

\end{thebibliography}
\end{document}